\documentstyle[aps,preprint,12pt,prb]{revtex}
\begin{document}
\draft
\title{Relation between Energy Level Statistics and Phase Transition
and its Application to the Anderson Model}
\author{E. Hofstetter and M. ${\rm Schreiber}^{*}$}
\address{Institut f\"{u}r Physikalische Chemie,
Johannes-Gutenberg-Universit\"{a}t, Jakob-Welder-Weg 11,\\
D-55099 Mainz, Federal Republic of Germany}
\address{$ ^{*}$ Fachbereich Physik, Technische Universit\"{a}t
Chemnitz-Zwickau, PSF 964, \\
D-09009 Chemnitz, Federal Republic of Germany}
\date{\today}
\maketitle
\begin{abstract}
A general method to describe a second-order phase transition is discussed. It
starts from the energy level statistics and uses of finite-size scaling. It is
applied to the metal-insulator transition (MIT) in the Anderson model of
localization, evaluating the cumulative level-spacing distribution as well as
the Dyson-Metha statistics. The critical disorder $W_{c}=16.5$ and the critical
exponent $\nu=1.34$ are computed.
\end{abstract}
\vspace{0.5cm}
\twocolumn
The Anderson model is well known to adequately describe some characteristic
aspects
of disordered systems. In three dimensions (3D) it exhibits a metal-insulator
transition (MIT). In spite of the simplicity of the Anderson Hamiltonian, the
critical exponent remains a controversial subject theoretically \cite{IL,FW}
as well as numerically. \cite{BK} Recently previous contradictions concerning
the numerical results could be removed \cite{EH} and a universal value for the
exponent established for several numerical data sets. But all these numerical
studies were done in the framework of the transfer matrix method (TMM).
\cite{AMK} Here we discuss another approach, no longer based on the TMM data,
to calculate critical parameters like the critical disorder $W_{c}$ and the
critical exponent $\nu$ and confirm the previous results. The approach is
fairly general and can be easily applied to other phase transitions. Only for
clarity of our arguments we formulate the subsequent derivation in terms of
the Anderson model of localization and its MIT.
\par
The MIT is expected to be a second order phase transition which can be
characterized by the correlation length $\xi_{\infty}$ of the order parameter
fluctuations. This length diverges with the critical exponent $\nu$ at the
critical point $W_{c}$,
\begin{equation}
\xi_{\infty}(W)=\zeta|W-W_{c}|^{-\nu}.
\end{equation}
Here $W$ is the critical parameter and $\zeta$ a proportionality constant.
In the usual TMM the correlation length $\xi_{M}$ of a finite system of size
$M$ is calculated and extrapolated to an infinite system size by a
finite-size scaling procedure. In the following a characteristic quantity
$\alpha_{M}$ is used from which the critical parameters are derived.
There is no need to compute $\xi_{M}$ and one has a large choice for the
definition of $\alpha_{M}$.
\par
The present paper extends previous studies \cite{BS,EH1} of the statistical
properties of the energy spectrum of the Anderson Hamiltonian on both sides of
the MIT. In a similar way the influence of the magnetic field on
the spectrum with \cite{YO} or without \cite{GM} disorder has been considered.
After unfolding the spectrum \cite{OB} its fluctuations are characterized by
means of the spacing distribution $P(s)$ and the Dyson-Metha statistics
$\Delta_{3}$. The first entity measures the level repulsion between
consecutive levels and the second one reflects the spectral rigidity. In terms
of these quantities the MIT, in a system of finite size $M$, is a continuous
transition from the Gaussian orthogonal ensemble (GOE) to the Poisson
ensemble (PE). These two limiting regimes, GOE and PE, were derived by
Alt'shuler {\it et al.} using diagram technique for conduction electrons
moving in a random impurity potential and tested numerically for a small
system ($M=5$). \cite{BA} The values of $P(s)$ and $\Delta_{3}$ are functions
of the size $M$ except at the critical point. In the thermodynamic limit this
critical point behaviour is expected to separate two different regimes: GOE
for the metallic side and PE for the insulating side. Similar results have
already been mentioned by Evangelou {\it et al.} \cite{SE} and studied in
more detail by Shklovskii {\it et al.}\cite{BS} but considering only $P(s)$.
We consider a yet undefined characteristic quantity $\alpha(M,W)$
which shall be chosen in such a way that:
\begin{equation}
\alpha(M,W) \stackrel{M \rightarrow \infty}{\longrightarrow}
\left\{
\begin{array}{ll}
\tilde{\alpha}_{Me} & \forall W<W_{c} \\
\tilde{\alpha}_{Cr} &  W=W_{c}\\
\tilde{\alpha}_{In} & \forall W>W_{c}
\end{array}
\right.
\end{equation}
where $\tilde{\alpha}_{Me}$, $\tilde{\alpha}_{In}$ and $\tilde{\alpha}_{Cr}$
are different constants for the metallic side, the insulating side and at the
critical point, respectively. With respect to the $W$ dependence $\alpha(M,W)$
is singular for $M=\infty$, but analytical otherwise because all the
singularities connected with the phase transition are rounded for a finite
size. As $\alpha(M,W)$ describes a system which exhibits a second order phase
transition with characteristic length $\xi_{\infty}$, Eq.(1) suggests
generally, following the
work of Br\'{e}zin, \cite{EB} a finite-size scaling behaviour for
$\alpha(M,W)$,
namely $\alpha (M,W)/\tilde{\alpha}=g(M/\xi_{\infty}(W))$ or
$\alpha (M,W)=f(M/\xi_{\infty}(W))$. But as $\alpha(M,W)$ is analytical for
finite $M$, $f$ can be expanded around the critical point as
$f(x) \stackrel{x \rightarrow 0}{=} a+bx^{1/\nu}+cx^{2/\nu}+\ldots$, yielding
\begin{equation}
\alpha (M,W) \simeq \alpha (M,W_{c})+ C |W-W_{c}| M^{1/\nu}
\end{equation}
for any quantity $\alpha$ which fulfills the condition (2) with a singularity
at $W_{c}$ for $M\rightarrow\infty$ and which obeys the scaling relation
$\alpha (M,W)=f(M/\xi_{\infty}(W))$. We note that one has to consider a
{\it finite-size} scaling \cite{MF} procedure so that the size of the system
is kept finite, suitable for numerical work, and not usual scaling \cite{DT}
for which an infinite system has to be considered.
\par
Before defining $\alpha(M,W)$ we recall how $P(s)$ and $\Delta_{3}$ were
computed. \cite{EH1}
The energy spectra were obtained from the Anderson Hamiltonian
$H=\sum_{i}\epsilon_{i} |i\rangle \langle i| +\sum_{i\neq j} V |i\rangle %
\langle j|$, where the sites {\it i} are distributed on a simple cubic lattice,
interactions with only the nearest neighbours are considered and $V$=1 defines
the energy scale. The site energy $\epsilon_{i}$ is described by a stochastic
variable given by a box distribution of width $W$ which represents the disorder
and is our critical parameter. The secular matrix of the Hamiltonian was
diagonalized in a straightforward way by means of the Lanczos algorithm for
systems of size $M^{3}$ from $13^{3}$ to $21^{3}$ and disorder $W$ from 2 to
40.
The number of different realizations of $\epsilon_{i}$ was chosen so that
$\sim 2 \cdot 10^{5}$ eigenvalues were obtained for every pair of
parameters $(M,W)$. Thus the full spectrum was computed 25 to 90 times. For
the subsequent investigations only half of
the spectrum around the band center is considered. \cite{EH1}
\par
To analyse the scaling of the statistical properties of the eigenvalue
spectrum, starting from $P(s)$ an evident choice for $\alpha$, which satisfies
the expected finite-size behaviour, would be $\alpha=\int_{0}^{\infty} P(s)
ds$.
But because $P(s)$ is normalized this makes no sense. A solution is provided by
the appearance of a fixed point at $s_{0}\simeq 2$ when one plots $P(s)$ for
different disorders $W$ and sizes $M$.\cite{BS,EH1} So one can choose
$\alpha_{P(s)}(M,W)=\int_{s_{0}}^{\infty} P(s) ds$, with
$s_{0} \simeq 2$ as already quoted. \cite{BS} An equivalent choice \cite{EH1}
would be $\alpha_{P(s)}(M,W)=\int_{0}^{s_{0}} P(s) ds$. While these choices
are formally possible it is much better from a numerical point of view to
consider the cumulative level-spacing distribution
$I(s)=\int_{0}^{s} P(s') ds'$ as plotted in Fig.\ref{1} instead of $P(s)$
because the most significant changes which occur in $P(s)$ for small $s$ are
emphasized and because $I(s)$ is much smoother.
Using $I(s)$ we choose
\begin{equation}
\alpha_{1}(M,W)= \frac{1}{s_{0}}\int_{0}^{s_{0}} I(s) ds \;.
\end{equation}
The results in Fig.\ref{2} clearly distinguish three different regions, in
which $\alpha_{1}$ increases, remains constant, or decreases with increasing
size $M$, respectively. This is consistent with the properties (2) of
$\alpha(M,W)$ required in order to apply the method. We can already determine
the critical disorder from Fig.\ref{2}: $\alpha_{1}(M,W)$ is size independent
for $W_{c}=16.25\pm 0.25$.
\par
In the next step we check whether $\alpha_{1}(M,W)$ can be expressed by a
scaling
function $f(M/\xi_{\infty}(W))$. To perform this scaling procedure with the
$\alpha_{1}$ data only, the range of $\alpha_{1}$ values for various $M$ at
any given disorder $W_{1}$ must overlap the range of $\alpha_{1}$ values for
various $M$ for at least one different disorder $W_{2}$. That would require to
compute $\alpha_{1}(M,W)$ for more $W$ values or larger $M$. This is beyond our
present computational means. Because of this
restriction we use the values of $\xi_{\infty}$ from previous calculations done
in the framework of the TMM.\cite{MO} Of course, this does not contradict the
statement in the introduction that the {\it method} is no longer based on the
TMM; only the present evaluation is. The TMM data are {\it not necessary}
for the calculation of $W_{c}$ and $\nu$ below but just for the
demonstration of the scaling. The result in Fig.\ref{3} confirms the scaling
hypothesis for $\alpha_{1}(M,W)$. The upper branch of the curve represents
$W>W_{c}$ and the lower $W<W_{c}$. The fluctuations in the scaling curve can be
explained by the facts that one uses a correlation length $\xi_{\infty}$
obtained from a different method, that some values of $\xi_{\infty}$
in the critical region had to be interpolated, and that the range of the
$\alpha_{1}$ values is very small. The biggest fluctuations are due to the
smallest $M$ which indicates that $M=13$ may be not large enough or that one
has to average over more realizations.
\par
In the final step we determine the critical exponent $\nu$. The tricky point
is
that the formula (3) is valid only in the vicinity of the critical point but
on the other hand one knows \cite{BK,EH} that close to the transition the
numerical inaccuracies are largest. The task is therefore to find an adequate
range for $|W-W_{c}|$ which satisfies these two constraints. This can be
controlled by a $\chi^{2}$-fitting \cite{WP} which measures the agreement
between the data and the "model" (3). To determine the "best" range we
minimize $\chi^{2}/N$, where $N$ is the number of data considered, as a
function of the range $|W-W_{c}|$ and obtain $\nu=1.37\pm 0.15$ for
$|W-W_{c}|\simeq 1$ in very good agreement with our previous result \cite{EH}
$\nu\simeq 1.34$ for the box distribution. Using $\alpha_{P(s)}$ but with even
fewer data around the critical point and for smaller samples, Shklovskii
{\it et al.} \cite{BS} estimated $\nu \simeq 1.5$ which is consistent with our
results.
\par
Considering as above $I(s)$, only a part of the information contained in the
spectrum is evaluated, namely the correlation between consecutive levels.
By the $\Delta_{3}$-statistics (see Fig.\ref{4}) the correlation
between nonconsecutive levels is taken into account, too. Now we choose
\begin{equation}
\alpha_{2}(M,W)=\frac{1}{30}\int_{0}^{30} \Delta_{3}(L) dL
\end{equation}
The results in Fig.\ref{5} show that the critical disorder can be located at
$W_{c}=16.5\pm 0.1$. The quality of these data is better than that of
$\alpha_{1}(M,W)$: the fixed point is well defined and the different $M$
dependencies described above for $\alpha_{1}$ can be seen more accurately for
$\alpha_{2}$. The scaling assumption for $\alpha_{2}(M,W)$ is tested in the
same way as that for $\alpha_{1}$ and confirmed in Fig.\ref{6}. For the
critical exponent, using Eq.(3) and the $\chi^{2}$ fitting, one obtains
$\nu=1.34\pm 0.10$, identical to the value previously found, \cite{EH} with
$|W-W_{c}|\simeq 1$. As supposed above these results are more accurate than
the ones provided by $P(s)$ or $I(s)$.
\par
In conclusion, the here employed method to study the MIT on the basis of
the statistical properties of the energy spectrum provides very good results
qualitatively as well as quantitatively. The obtained critical exponent
$\nu=1.34 \pm 0.10$ confirms the previous \cite{EH} TMM result. Applying
the same TMM but a slightly different way to compute the critical exponent
from the raw data, MacKinnon \cite{AMK1} finds $\nu\simeq1.53$. While that
result is not in contradiction with our uncertainty, we believe that the
difference is significant. Theoretically the derivation of the critical
exponent remains an open subject although recently Hikami,\cite{SH} using
string theory, proposed the possibility to obtain $\nu=4/3$ which would be
supported by our results. The discrepancy between the numerical predictions
for $\nu$ and the experimental values, \cite{GT} between 0.5 and 1, raised
the question whether the Anderson model is relevant to describe disordered
systems. But recently experimental measurements of $\nu$ in uncompensated
Si:P samples have yielded $\nu=1.3$.\cite{HST} So while it cannot yet be
claimed that the problem of the critical exponent in disordered systems is
solved it seems that results obtained in very different ways converge towards
the same value $\nu\simeq 1.35$.
\par
It is surprising that the same value has
been found applying the TMM to 3D systems with magnetic field. \cite{MH} The
analysis of the statistical properties of the energy spectrum can be a
promising way to understand this. Until now it was claimed that one could
expect different critical exponents for different universality classes
e.g. the orthogonal (GOE) and the unitary (GUE) classes. However, new results
have shown \cite{BS,EH1} this type of symmetry only for small disorder but not
at the critical point where one finds an intermediate distribution e.g.,
between GOE and PE in the orthogonal case. The same critical exponent should
mean the same universality class {\it at} the critical point. Thus the
observed agreement of the critical exponents with and without magnetic field
may be interpreted as a signature of a new universality class different from
PE, GOE, and GUE. \cite{EH2}
\par
Finally we stress that although the method was employed for the MIT in the
Anderson model of localization, the approach is very general and easy
to apply. It would be very interesting to study other second order phase
transitions from this point of view not necessarily considering the eigenvalues
of the Hamiltonian, but for example the eigenvalues of the transfer matrix
in Ising models.

\newpage

\begin{figure}
\caption{Cumulative level-spacing distribution $I(s)$ for a system size
$21 \times 21 \times 21$ and the indicated values of the disorder $W$.
The solid lines represent $I(s)$ for the GOE and the PE cases.}
\label{1}
\end{figure}

\begin{figure}
\caption{Scaling variable $\alpha_{1}(M,W)$ derived from $I(s)$, according
to Eq.(4). For the GOE and the PE the limiting values $\tilde{\alpha}_{Me}
\simeq 0.507$ and $\tilde{\alpha}_{In}\simeq 0.569$, respectively, can be
derived.}
\label{2}
\end{figure}

\begin{figure}
\caption{Verification of the scaling assumption for $\alpha_{1}(M,W)$.
The correlation length $\xi_{\infty}$ is taken from
TMM data. \protect \cite{MO}}
\label{3}
\end{figure}

\begin{figure}
\caption{$\Delta_{3}$-statistics for a system size $21 \times 21 \times 21$
 and the values of the disorder $W$ indicated on the right hand side.}
\label{4}
\end{figure}

\begin{figure}
\caption{Scaling variable $\alpha_{2}(M,W)$ derived from $\Delta_{3}(L)$
according to Eq.(5). For the GOE and the PE the limiting values
$\tilde{\alpha}_{Me}\simeq 0.264$ and $\tilde{\alpha}_{In}\simeq 0.804$,
respectively, can be derived.}
\label{5}
\end{figure}

\begin{figure}
\caption{Verification of the scaling assumption for $\alpha_{2}(M,W)$.
The correlation length $\xi_{\infty}$ is taken from
TMM data. \protect \cite{MO}}
\label{6}
\end{figure}


\begin{references}
\bibitem{IL}
I. V. Lerner, in {\it Localisation 1990}, eds. K. A. Benedict and
J. T. Chalker, Institute of Physics Conf. Ser. {\bf 108}, 53 (1991).
\bibitem{FW}
F. Wegner, Nucl. Phys. B {\bf 316}, 663 (1989).
\bibitem{BK}
B. Kramer {\it et al.}, Physica A {\bf 167},
163 (1990).
\bibitem{EH}
E. Hofstetter and M. Schreiber, Europhys. Lett. {\bf 21}, 933 (1993).
\bibitem{AMK}
A. MacKinnon and B. Kramer, Phys. Rev. Lett. {\bf 47}, 1546 (1981);
Z. Phys. B {\bf 53}, 1 (1983).
\bibitem{BS}
B. I. Shklovskii {\it et al.}, Phys. Rev. B {\bf 47}, 11487 (1993).
\bibitem{EH1}
E. Hofstetter and M. Schreiber, Phys. Rev. B {\bf48}, 16979 (1993).
\bibitem{YO}
Y. Ono {\it et al.}, J. Phys. Soc. Jpn.
{\bf 62}, 2762 (1993).
\bibitem{GM}
N. Dupuis and G. Montambaux, Phys. Rev. B {\bf 43}, 14390 (1991).
\bibitem{OB}
O. Bohigias and J.-M. Giannoni, in {\it Mathematical and Computational
Methods in Nuclear Physics}, eds. J. Dehesa, J. Gomez and A. Polls,
Lecture Notes in Physics {\bf 209}, 1 (Springer, Berlin, 1984);
O. Bohigias {\it et al.}, Phys. Rev. Lett. {\bf 52},
1 (1984).
\bibitem{BA}
B. L. Alt'shuler and B. I. Shklovskii, Sov. Phys. JETP {\bf 64}, 127 (1986);
B. L. Alt'shuler {\it et al.}, Sov. Phys. JETP {\bf 67}, 625 (1988)
\bibitem{SE}
S. N. Evangelou and E. N. Economou, Phys. Rev. Lett. {\bf 68}, 136 (1992).
\bibitem{EB}
E. Br\'{e}zin, J. Physique {\bf 43}, 15 (1982).
\bibitem{MF}
M. E. Fisher and M. N. Barber, Phys. Rev. Lett. {\bf 28}, 1516 (1972).
\bibitem{DT}
D. J. Thouless, Phys. Rev. Lett. {\bf 39}, 1167 (1977).
\bibitem{MO}
M. Ottomeier, unpublished.
\bibitem{WP}
W. H. Press {\it et al.}, {\it Numerical Recipes}
(University Press Cambridge, 1989).
\bibitem{AMK1}
A. MacKinnon, to be published.
\bibitem{SH}
S. Hikami, Prog. Theor. Phys. Suppl. {\bf 107}, 213 (1992).
\bibitem{GT}
G. A. Thomas and M. A. Paalanen, in {\it Localization, Interaction and
Transport Phenomena}, eds. B. Kramer, G. Bergmann and Y. Bruynseraede,
Springer Series in Solid State Sciences {\bf 61}, 77 (1985).
\bibitem{HST}
H. Stupp {\it et al.}, Phys. Rev. Lett. {\bf 71}, 2634 (1993).
\bibitem{MH}
T. Ohtsuki {\it et al.}, Solid State Commun.  {\bf 81}, 477 (1992);
M. Henneke {\it et al.}, Europhys. Lett. , submitted.
\bibitem{EH2}
E. Hofstetter and M. Schreiber, to be published.
\end{references}
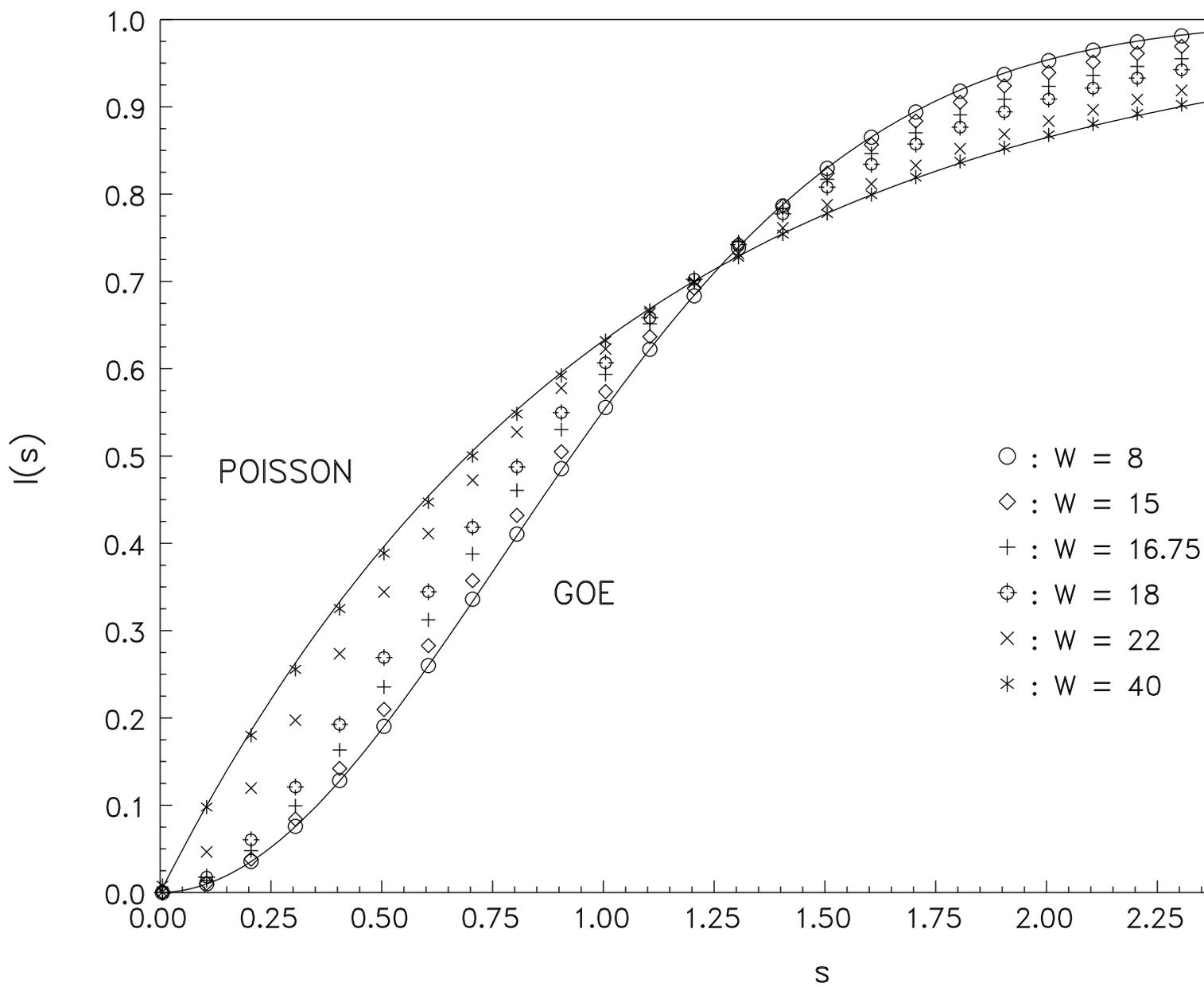
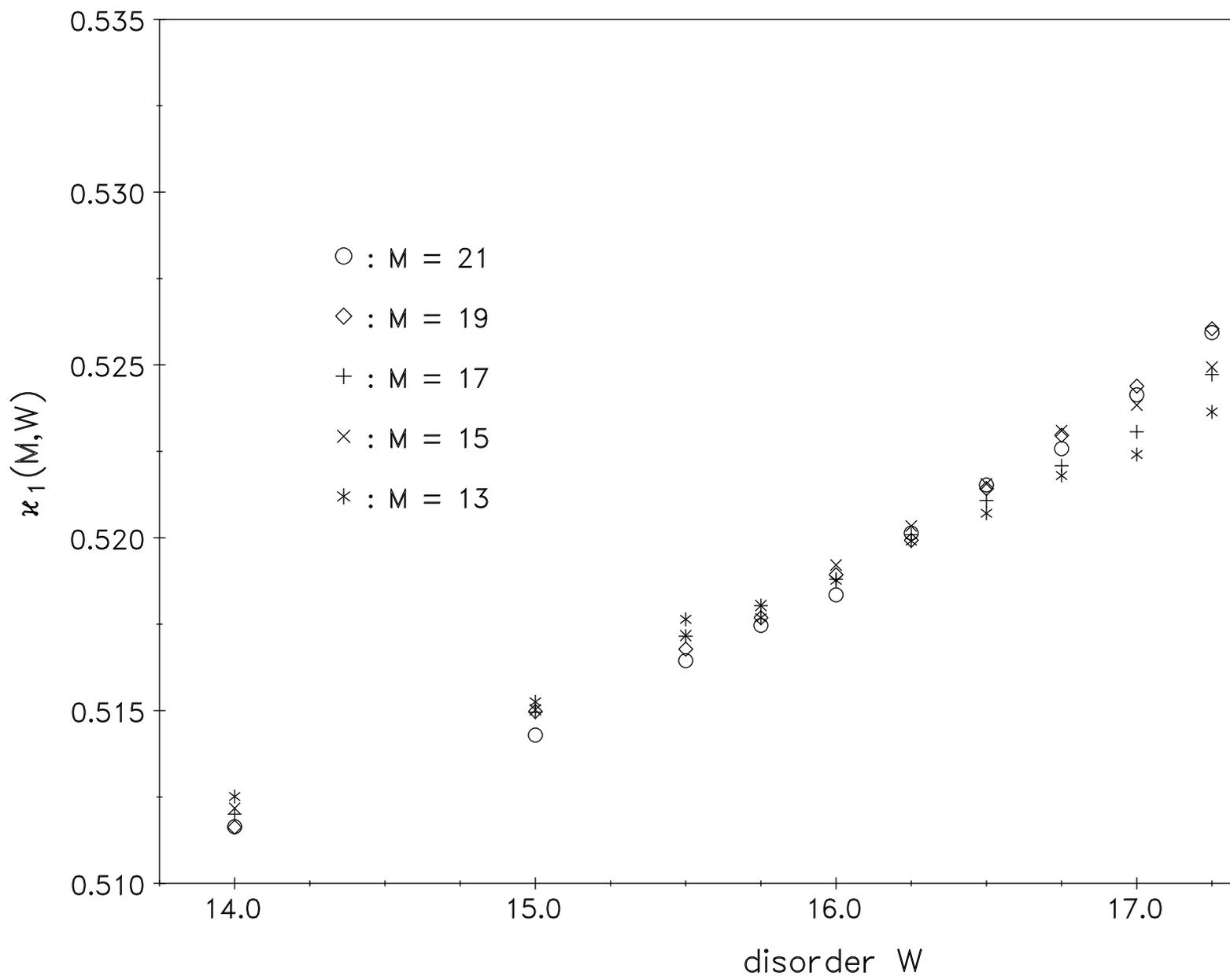
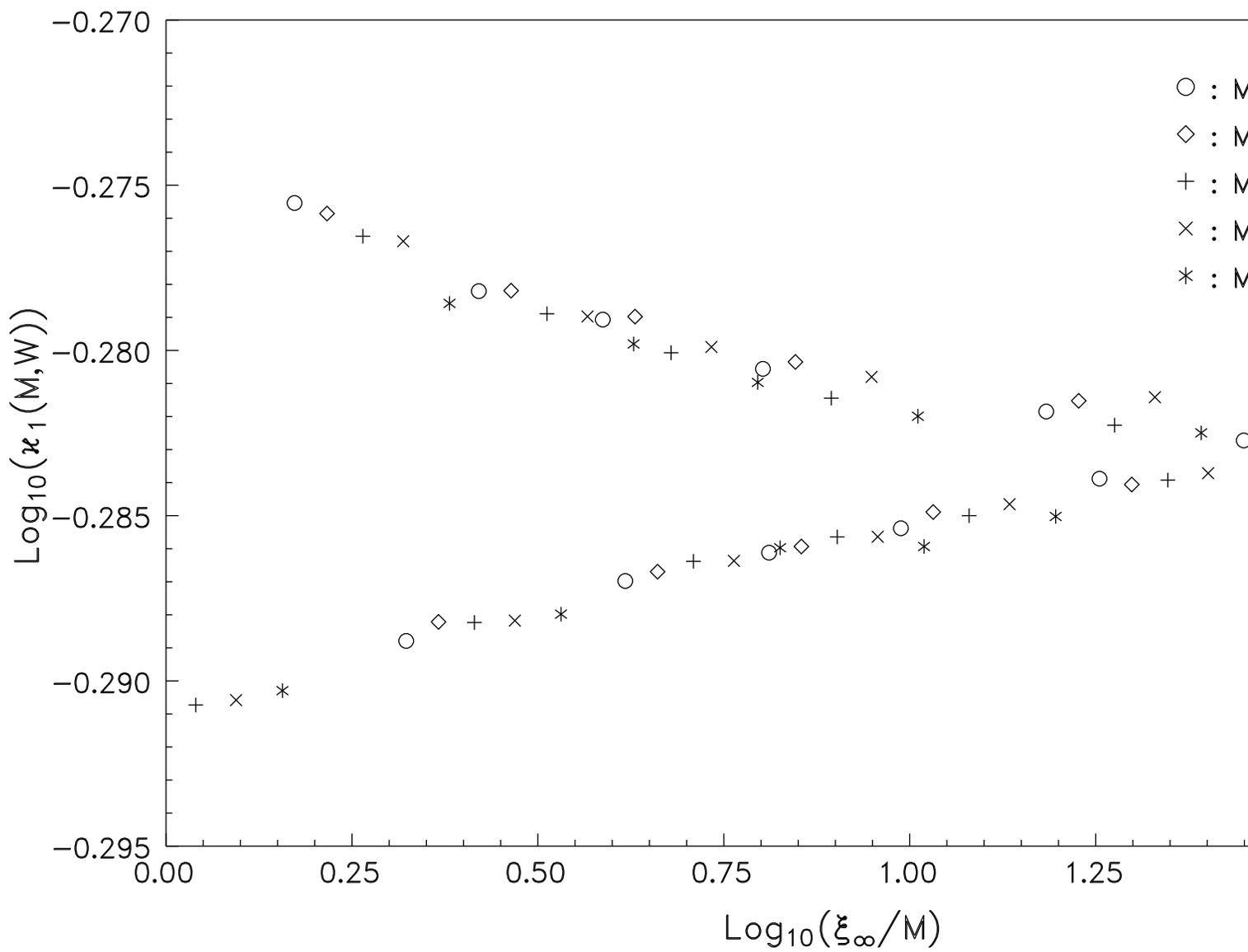
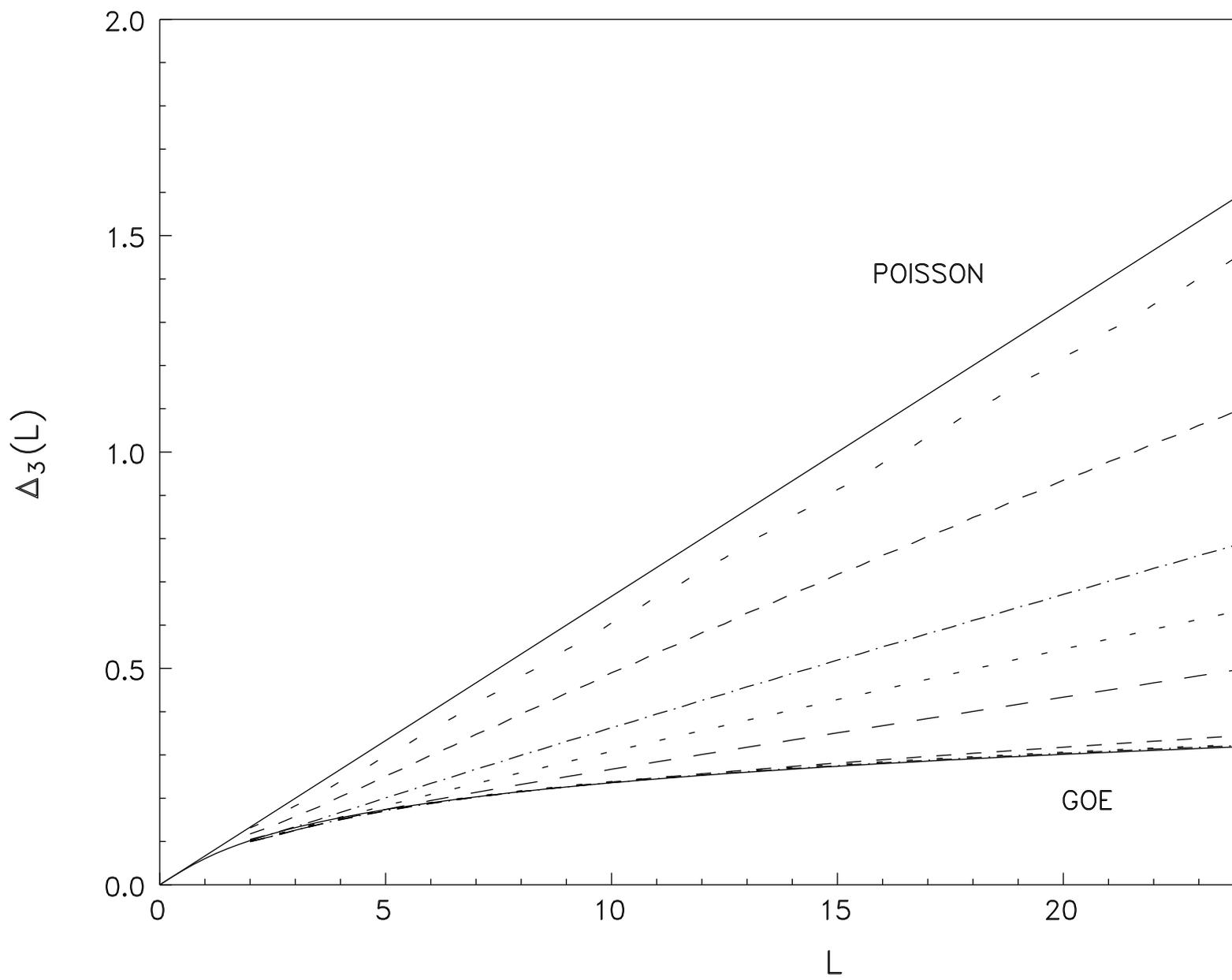
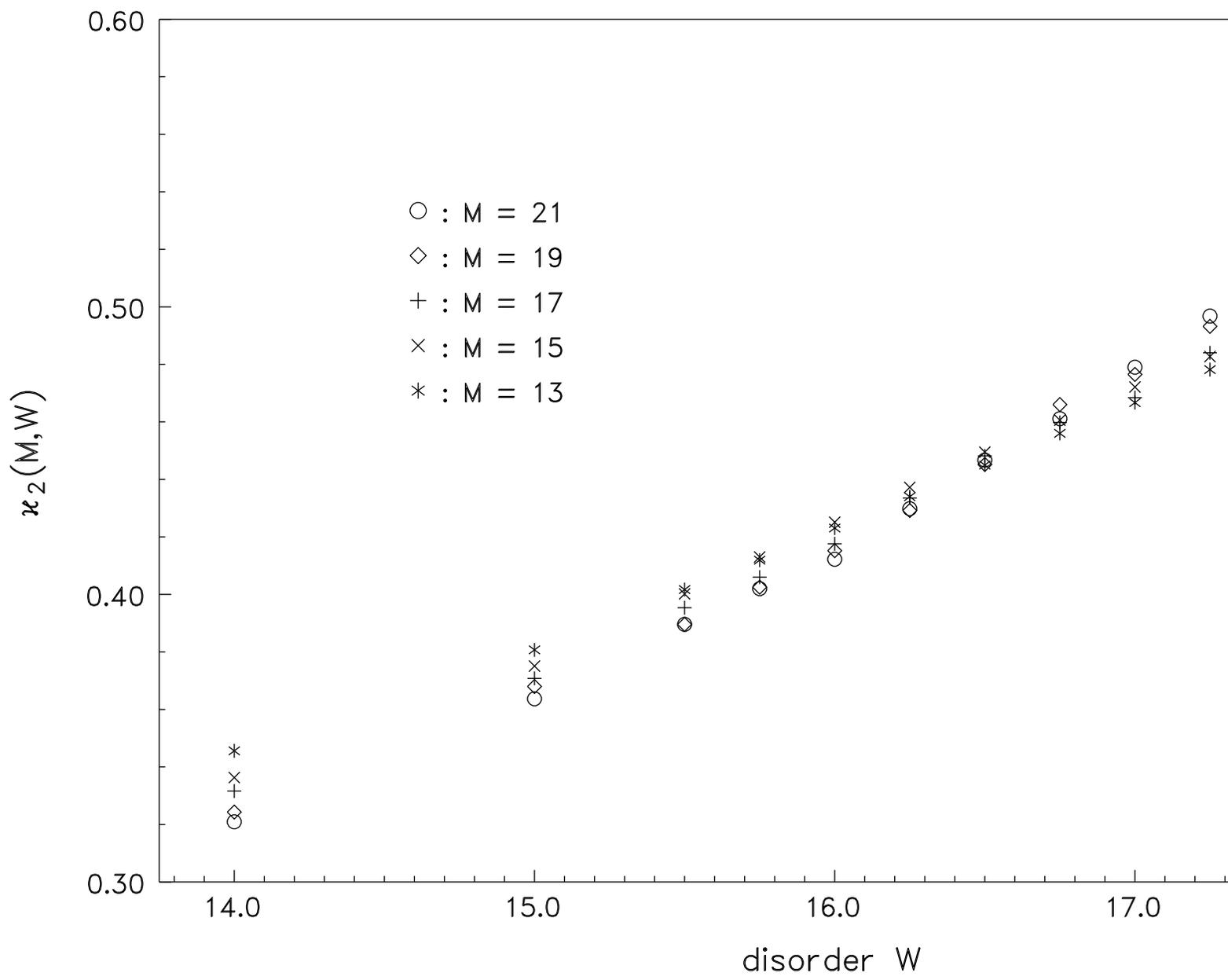
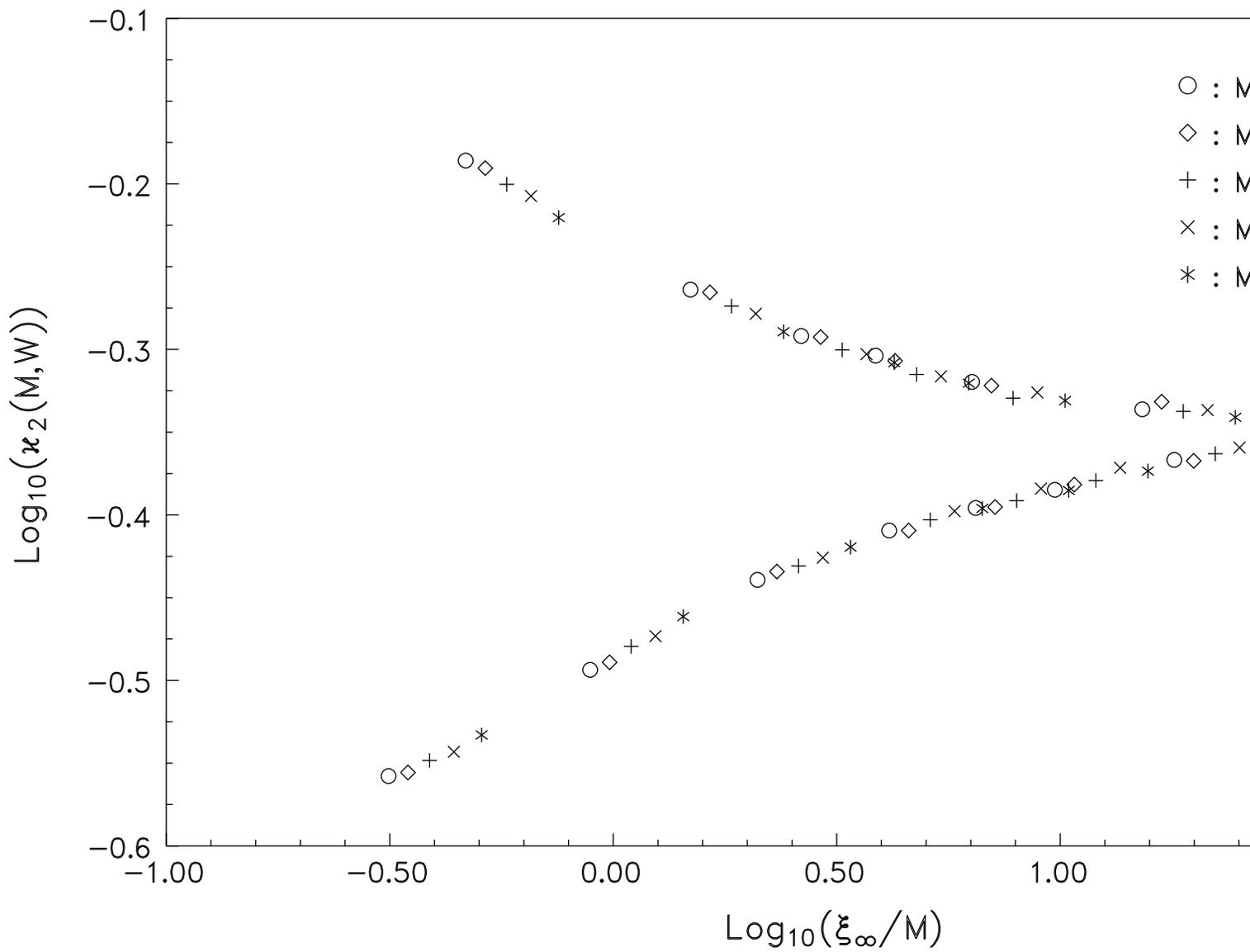
\end{document}